\documentstyle[12pt,aaspp4]{article}
\begin{document}
\parindent=1.0cm

\title{THE GLOBULAR CLUSTER CONTENT OF MAFFEI 1}

\author{T. J. Davidge \footnote[1]{Visiting Astronomer, Canada-France-Hawaii Telescope, 
which is operated by the National Research Council of Canada, the Centre National de le 
Recherche Scientifique, and the University of Hawaii.}}
\affil{Canadian Gemini Office, Herzberg Institute of Astrophysics,
\\National Research Council of Canada, 5071 West Saanich Road,
\\Victoria, B.C. Canada V9E 2E7\\ {\it email: tim.davidge@nrc.ca}}

\author{Sidney van den Bergh}
\affil{Herzberg Institute of Astrophysics,
\\National Research Council of Canada, 5071 West Saanich Road,
\\Victoria, B.C. Canada V9E 2E7\\ {\it email: sidney.vandenbergh@nrc.ca}}

\begin{abstract}

	Near-infrared images have been used to search for bright globular clusters 
near the center of the elliptical galaxy Maffei 1. The number of objects in various 
radial intervals from the center of Maffei 1 are compared with the number density of 
sources in two control fields, and we find an excess of $31 \pm 11$ objects with $K$ 
between 16.6 and 18.0 in an annulus between 20 and 90 arcsec from the galaxy 
center. At radii in excess of 90 arcsec the ability to find clusters is 
frustrated by statistical noise in the foreground star population. It is demonstrated 
that the globular clusters located within 90 arcsec of the center 
of Maffei 1 span a range of near-infrared colors that are at least as great 
as that which is seen among M31 clusters; some of the clusters appear to have 
colors that are consistent with them being very young. The specific globular cluster 
frequency within 90 arcsec of the center of Maffei 1 is S$_N = 1.3 \pm 0.6$ if the 
distance modulus is $28.1 \pm 0.2$ and A$_V = 4.7 \pm 0.2$. The specific cluster frequency 
of Maffei 1 is thus similar to that of other nearby field elliptical galaxies. 

\end{abstract}


\section{INTRODUCTION}

	With typical masses near $10^5$ M$_{\odot}$, globular clusters likely formed in 
molecular complexes that are much larger than those in the solar neighborhood, where the 
vast majority of molecular clouds have log(M$_\odot) < 5 - 6$ (e.g. Solomon \& Rivolo 
1989). As noted by Harris \& Pudritz (1994), the conditions required for the supergiant 
molecular clouds (SGMCs) that are likely globular cluster nurseries to form in 
large numbers are associated with significant events in the evolution of a galaxy, such 
as the initial collapse of protogalactic fragments and major mergers, both of which spur 
the cloud-cloud collisions that are thought to be needed for the assembly of large 
star-forming complexes. Events of this nature can occur at various times in the life of a 
galaxy; consequently, the clusters in a galaxy may not be coeval, but may have formed 
during different times and under different conditions. Indeed, many elliptical 
galaxies have red and blue globular clusters, and the absence of a correlation 
between the metallicity of blue globular clusters and host galaxy properties (e.g. 
Forbes \& Forte 2001, but see also Strader, Brodie, \& Forbes 2004) suggests that these 
objects formed very early on, possibly in autonomous mini-halos (e.g. Beasley et al. 
2003). On the other hand, there are indications that red globular clusters may have 
formed after the potential well of the final galaxy had been defined, possibly during 
mergers (e.g. Kundu \& Whitmore 2001). Some objects that 
appear to be globular clusters at the present day may actually be the cores of 
disrupted satellites (e.g. van den Bergh \& Mackey 2004; Bekki \& Freeman 
2003; Mizutani, Chiba, \& Sakamoto 2003; Meylan et al. 2001).

	The events that lead to the formation of 
SGMCs are believed to be associated with the creation of 
pressure-supported, as opposed to rotationally supported, structures. It 
might thus be anticipated that the specific frequency of globular clusters, S$_N$ (Harris 
\& van den Bergh 1981), should be higher in galaxies dominated by spheroids, rather than 
disks. This expectation is broadly consistent with observations, although there 
also appears to be an environmental dependence, in the sense that S$_N$ tends to be 
highest among elliptical galaxies near the centers of rich clusters (e.g. Harris 1991, 
but see also Kundu \& Whitmore 2001). While it has been suggested that some Galactic 
globular clusters may be associated with the disk (e.g. Zinn 1985), these 
more likely formed with the bulge (e.g. Minniti 1995).

	Studies of the globular clusters systems around nearby galaxies provide an 
obvious means of testing our understanding of the formation and evolution of globular 
clusters. As one of the nearest elliptical galaxies, Maffei 1 is a potentially important 
target for a survey of globular cluster content. A complicating factor is that Maffei 1 
is viewed through the Galactic disk, and so there is significant obscuration by dust and 
contamination from foreground stars. The extinction towards Maffei 1 is patchy, and 
individual dense dust clouds are seen projected 
against the central regions of the galaxy (e.g. Buta \& McCall 2003).

	Maffei 1 might be expected to have a large number of clusters. 
Indeed, elliptical galaxies have a mean specific cluster frequency S$_N = 
2.4 \pm 0.8$ (Kundu \& Whitmore 2001). Therefore, with $V = 11.1$ (Buta \& McCall 1999), 
so that M$_V = -21.7$, Maffei 1 should have an entourage of $\sim 1150$ clusters if 
it is typical of other elliptical galaxies. However, only a modest number of 
clusters have been detected to date. Davidge (2002) used $J, H,$ and $K'$ images obtained 
with the CFHT AOB $+$ KIR imager to search for globular clusters in the central $700 
\times 700$ parsec$^2$ of this galaxy, and no obvious cluster candidates brighter than 
the peak of the globular cluster luminosity function (GCLF) were detected. Using WFPC2 
images, Buta \& McCall (2003) identified 20 non-stellar objects that have sizes and 
brightnesses which are consistent with them being 
clusters. However, the mean $R-I$ colors of these objects, 
as well as the scatter about the mean, differ from those of the Galactic globular cluster 
system. It is not clear if these differences in photometric properties are intrinsic 
to clusters in Maffei 1, or if they are a consequence of differential reddening.

	In the present study, broad-band near-infrared images are used to search for 
globular clusters in the central $6 \times 6$ arcmin$^2$ ($7.3 \times 7.3$ kpc$^2$) of 
Maffei 1. This area contains slightly more than half of the total light from Maffei 1 
(Buta \& McCall 1999), and so should also contain a significant fraction of the Maffei 1 
cluster system. A search for globular clusters in Maffei 1 at near-infrared wavelengths 
has two advantages when compared with searches at visible wavelengths. First, the 
extinction is much smaller in the infrared, with A$_K$/A$_V \sim 0.11$ (Rieke \& Lebofsky 
1985). Second, the broad-band near-infrared spectral-energy 
distributions (SEDs) of stars are sensitive to surface gravity, thereby simplifying 
the task of distinguishing between the low surface gravity giants that dominate the 
integrated spectra of globular clusters and the high surface gravity main sequence stars 
that occur in large numbers in the foreground disk. This behaviour is clearly seen in the 
$(J-H, H-K)$ two-color diagram (TCD), which is used to investigate the SEDs of 
sources towards Maffei 1 in this paper.

	Reddening and distance estimates for Maffei 1 are summarized in Tables 1 and 2. 
For the current study we select the median value in each table, and so a distance 
modulus of $28.1 \pm 0.2$ and a foreground absorption A$_V = 4.7 \pm 0.2$ magnitudes are 
adopted throughout the paper. 

	The paper is structured as follows. Details of the observations and the data 
reduction procedures are reported in \S 2, while the photometric measurements 
and the results of the search for globular clusters are presented in 
\S 3. The S$_N$ of the central regions of Maffei 1 is estimated in \S 4, while 
a discussion and summary of the results follows in \S 5.

\section{OBSERVATIONS \& REDUCTIONS}

	The data were recorded during the nights of November 21 and 22 (UT)
2002 with the CFHTIR imager, which was mounted at the Cassegrain focus of 
the 3.6 metre Canada-France-Hawaii Telescope (CFHT). The detector in the CFHTIR is 
a $1024 \times 1024$ HgCdTe array. Each pixel subtends 0.22 arcsec 
on a side, and so individual exposures cover a $3.8 \times 3.8$ arcmin$^2$ area.

	The central $6 \times 6$ arcmin$^2$ of Maffei 1 was imaged by recording 
exposures at four different pointings. Images were recorded through $J, H,$ and $K'$ 
filters with a total exposure time of 12 minutes per filter per pointing. 
The layout of the pointings on the sky 
is shown in Figure 1. Also shown in Figure 1 is the $\mu_I = 22.77$ magnitude 
arcsec$^{-2}$ isophote, which encompasses half of the total light from Maffei 1 and 
was computed using the structural information plotted in Figures 23 and 24 of Buta 
\& McCall (1999). Adjacent pointings overlap, and sources in the regions that are common 
to adjacent pointings were used to assess photometric stability (\S 3.1). 

	Contamination by foreground stars is a major consideration when observing 
Maffei 1. Two control fields, located 30 arcminutes to the 
north and south of the galaxy center, were observed with the same filters and 
exposure times as Maffei 1 to monitor foreground star contamination. The co-ordinates of 
the control fields are listed in Table 3.

	Calibration images were acquired at various times during the three night 
observing run. Dome flats were recorded at the beginning of each night, while the 
detector dark current was measured at the end of each night. Neither the dark current 
nor flat field pattern showed night-to-night variations, and so the relevant data from 
individual nights were combined to construct master dark and flat-field frames for the 
entire run. Regions of the sky with low stellar surface density were also observed 
periodically, and these were used to construct calibration images that monitor 
interference fringes and thermal emission from warm objects along the optical path. 

	The data were processed using a standard sequence for infrared imaging
that removes signatures introduced by the telescope, 
the instrumentation, and the atmosphere. The basic steps 
in this sequence are (1) the subtraction of a dark frame, (2) the 
division by a flat-field frame, (3) the subtraction of the DC sky level from 
each exposure, and (4) the subtraction of a fringe/thermal emission calibration 
frame. The processed images for each field $+$ filter combination were 
spatially registered to correct for dither offsets made during 
acquisition and then median combined. The resulting combined images were 
trimmed to the area common to all exposures of that pointing. The 
seeing was variable when these data were recorded. It was therefore decided not to 
construct a mosaic of the entire imaged region, but to consider the 
individual pointings seperately for the photometric analysis.
Examples of the final images are shown in Figures 2 and 3, which show the final $K'$ 
images of Pointing 4 and Control Field 2, respectively.
The image quality of the Maffei 1 and control field data are mediocre by 
Mauna Kea standards, and stars in the final images have FWHM between 0.9 and 1.1 arcsec.

\section{RESULTS}

\subsection{Photometric Measurements}

	Unresolved stars in Maffei 1 form a nonuniform background that introduces 
uncertainties in the photometric measurements, especially those of objects 
within roughly 30 arcseconds of the galactic nucleus. This background galaxy light 
was removed by applying a $10 \times 10$ arcsec$^{2}$ median boxcar filter 
to the processed images to suppress point sources; the result was subtracted 
from the unsmoothed data. The brightnesses of individual stars 
were then measured from the background-subtracted data with the point 
spread function (PSF) fitting routine ALLSTAR (Stetson \& Harris 1988), using 
co-ordinates, preliminary stellar brightnesses, and PSFs obtained from 
routines in the DAOPHOT (Stetson 1987) package. 

	The photometric calibration was defined using standard stars from Hawarden et al. 
(2001), which were observed throughout the three night observing run. The standard 
deviation in the photometric zeropoints obtained from standards stars that were 
observed during the first night is $\pm 0.1$ mag in each bandpass, suggesting that thin 
cirrus may have been present on that night, which is when data for Pointings 1 and 2 were 
obtained. For comparison, the standard deviation of the standard star measurements 
on the second and third nights are $\pm 0.06$ magnitudes, indicating 
better photometric stability than during the first night.

	The brightnesses of sources in the overlapping portions of Pointings 1 and 2 were 
compared to check if clouds were present when Maffei 1 was observed during the first 
night, and the mean differences between the brightness measurements of sources 
with $J$, $H$, and $K$ between 16 and 17 are shown in Table 4. The 
uncertainties in this table are the standard errors in the mean. Also listed are 
the mean differences between the brightnesses for sources in the overlapping regions of 
Pointings 3 and 4, which were observed on the second night. The entries in this table 
indicate that the pointing-to-pointing photometric stability on each night was within a 
few hundredths of a magnitude; hence, if thin cirrus was present, then it was uniformly 
distributed and stable while the observations were recorded. Nevertheless, 
a comparison between the brightnesses of sources in the overlapping portions 
of Pointings 1 and 3 and Pointings 2 and 4 indicates that there is a slight offset 
between the nightly photometric calibrations. This was greatest in $H$ with 
$\Delta H = -0.122 \pm 0.012$, while for $J$ and $K$ the differences were $-0.076 \pm 
0.012$ and $-0.049 \pm 0.017$. To adjust for this, the brightnesses of sources in 
Pointings 1 and 2 were shifted by the mean differences to agree with the calibration of 
Pointings 3 and 4.

	Artificial star experiments were run to estimate sample 
completeness. The artificial stars were assigned colors 
appropriate for a `typical' M31 globular cluster, as discussed in \S 3.4, after applying 
an extinction correction A$_V = 4.7$ (\S 1). Following the procedures to identify and 
characterize individual cluster candidates that are discussed in \S 3, an artificial star 
was considered to be recovered only if it was detected in all three bandpasses. 

	The elevated background levels near the center of Maffei 1 cause incompleteness 
to become significant at a brighter magnitude than is the case at larger 
galactocentric distances. To account for the spatially nonuniform completeness function, 
each pointing was divided into a $3 \times 3$ grid, and the incompleteness in each cell 
was computed separately. The completeness map for stars with $K$ between 17.75 and 18.25 
in Pointing 4 is shown in the upper panel of Figure 4. The completeness fraction, $C$, is 
lowest in Cell 1, which contains the center of Maffei 1, while $C$ 
is more or less uniform over the remainder of the field. 

	The behaviour of $C$ with $K-$band magnitude in three cells that cut 
diagnonally across Pointing 4 is shown in the lower panel of Figure 4, and it is 
evident that $C$ plummets outside of Cell 1 when $K > 18$. Therefore, $K = 18$ is 
adopted as the faint limit when searching for globular clusters around Maffei 1. 
The completeness functions of Pointings 2 and 3 show similar behaviour, 
while incompleteness sets in at brighter magnitudes in Pointing 1 (\S 3.2).

\subsection{Near-infrared CMDs of the Central Region of Maffei 1}

	The $(K, H-K)$ and $(K, J-K)$ CMDs of the Maffei 1 
and control fields are compared in Figures 5 and 6. The CMDs of the galaxy 
and control fields are very similar, indicating that the majority of objects 
near the center of Maffei 1 with $K < 18$ are foreground stars. This foreshadows one of 
the main results of this paper, which is that globular clusters near the center of Maffei 
1 make only a minor contribution to the overall source counts in this part of the sky. 
The $H-K$ and $J-K$ colors of the foreground star sequence also show good field-to-field 
stability, indicating that large reddening differences on arcminute to degree scales are 
abscent. It thus appears that the dense dust clouds found in the vicinity of Maffei 1 
by Buta \& McCall (1999; 2003) do not affect greatly the near-infrared colors 
of objects with $K < 18$.

	Why do the near-infrared colors of objects with $K < 18$ show no evidence of the 
patchy dust absorption that is seen towards Maffei 1? To be sure, the uniform colors are 
partly due to the decreased sensitivity to extinction in the near-infrared. However, a 
more important factor is that the majority of sources with $K < 18$ along this sight line 
are probably in front of much of the obscuring material. Buta \& McCall (1999, 2003) 
discuss the origin of the dust that is projected against the inner regions of Maffei 
1. Buta \& McCall (1999) point out that many elliptical galaxies have obscuring dust 
lanes, and so it would not be surprising if this material were ultimately found to 
belong to Maffei 1. However, Buta \& McCall (2003) argue that the dust is likely in the 
foreground disk. If this is the case then most of the dust will probably be in the Perseus 
Arm of the Milky-Way, through which Maffei 1 is viewed. Many of the open clusters within 
a few degrees of Maffei 1 tend to have distances, based on the entries in the 
catalogue of Dias et al. (2002), between 2 and 4 kpc, and this is adopted in the 
following discussion as the distance to the Perseus Arm. 

	Based on the stellar content of the Solar Neighborhood, it can be anticipated that 
most of the stars along the line of sight will be M dwarfs, and stars of this 
spectral type that have $K < 18$ will be located on the near side of the Perseus Arm 
or in the interarm region -- i.e. in front of the region expected to contain the bulk 
of the dust that is obscuring Maffei 1. For example, an M0 V star (M$_K = +5.2$; Cox 
2000) with $K = 17$ will be at a distance of 2.3 kpc if there is no extinction, and will 
be even closer if some extinction is present. M dwarfs with later spectral types that are 
observed at the same $K$ magnitude will have even smaller distances, as they are 
intrinsically fainter; for example, an M5 V dwarf (M$_K = 6.1$) with $K = 17$ will be at a 
distance of at most 1.5 kpc. Thus, the majority of stars with $K < 18$ along this sight 
line will likely not be obscured by material in the Perseus Arm. This helps to explain 
the uniformity in the colors of the stellar sequences, as the 
localised areas of high extinction along this line of sight are more 
distant than the majority of stars.

	The CMDs of Pointing 1 are peculiar, in that the foreground star 
sequence truncates near $K = 17.5$. The agreement with the photometric measurements in 
the overlapping portions of Pointing 2 and 3, discussed in \S 3.1, indicates that 
the elevated faint limit in Pointing 1 is not due to clouds. While  an extremely thick 
foreground dust cloud could cause a truncation in the CMD of Pointing 1, 
most of the dust clouds in the vicinity of Maffei 1 
are located in the northern part of the galaxy (Buta \& McCall 1999, 2003, but see also 
Ford \& Jenner 1971, who trace a dust lane with a width of $\sim 10$ arcsec through 
part of Pointing 1). 

	The elevated faint limit in Pointing 1 is due to a higher than normal detector 
noise level, which was confirmed by measuring the standard deviation of the counts 
in a portion of the processed Pointing 1 $K-$band image that is devoid of objects. The 
problem appears to be restricted to the Pointing 1 data, and the artificial star 
experiments confirm that the completeness fraction of the 
Pointing 1 data are markedly lower when $K > 17$ 
than in the other pointings. While the Pointing 1 data will not be completely discarded, 
they are of limited use for the globular cluster search, and so will not be considered 
when computing the specific cluster frequency in \S 4.

\subsection{The Identification of a Centrally Concentrated Population of Globular Clusters}

	We use the $K$ LFs of the various fields to investigate the field-to-field 
uniformity of the number counts and conduct an initial reconnaisance 
for a population of globular clusters, which will show up as an excess population of 
objects with respect to the source counts in the control fields. The LFs 
of the various fields, corrected for incompleteness, are compared in Figure 7. 
The number of objects in the two control fields, which are 
separated by a degree on the sky, are remarkably similar, indicating that field-to-field 
variations in the density of foreground stars are abscent at infrared wavelengths when 
$K < 18$. Therefore, the source counts in these fields can be averaged to reduce noise. 
This uniformity is also seen in the number counts of the Maffei 1 data that, with the 
exception of Pointing 1, are remarkably similar over the entire range of brightnesses.

	An excess population of objects is present near $K = 18$ in Pointings 3 
and 4, although the significance of this excess is modest. The statistical 
significance of the excess population can be enhanced by examining regions that are close 
to the center of Maffei 1, which is where the density of clusters might be 
expected to be highest. We further refine the search by restricting the investigation to 
the magnitude range expected for globular clusters in Maffei 1. Barmby et 
al. (2001) find that the GCLF of M31 peaks near M$_K = -10$, and the LF shown in their 
Figure 2 indicates that over 90\% of the clusters occur within $\pm 2$ magnitudes 
of the peak brightness. If the GCLF of Maffei 1 is like that in M31 
then the peak should occur near $K = 18.6$ with the adopted distance and reddening, 
and the vast majority of clusters should have $K > 16.6$. Incompleteness is significant 
in our data when $K > 18$ (\S 3.1), and so the CFHTIR data do not sample the peak of 
the GCLF, unless the adopted distance modulus is greatly in error.

	While it is desireable to extend the search for clusters into the innermost 
regions of Maffei 1, where the contrast with respect to the foreground stars might 
be highest, the bright unresolved body of Maffei 1 greatly elevates the noise level near 
the galaxy center, and this consideration sets the innermost radius for our cluster 
search. A visual inspection of the images indicate that when $r < 20$ arcsec the data 
are affected by noise flucuations, and so the inner radius for the cluster search was 
fixed at 20 arcsec. The artificial star experiments indicate that the brightness of 
sources with $K = 18$ are not affected by noise when $r > 20$ arcsec.

	The number of objects with $K$ between 16.6 and 18.0 were counted in four radial 
intervals centered on the galaxy nucleus: 20 -- 30 arcsec, 30 -- 60 arcsec, 60 -- 90 
arcsec, and $90 - 120$ arcsec. The raw counts were corrected for incompleteness using 
the results from the artificial star experiments. The total number of 
objects in each radial interval in Pointings 2, 3, and 4 
are shown in the second column of Table 5. The number of 
sources expected in each radial interval based on the control field data is shown in the 
third column of Table 5; the uncertainties in these entries are small because 
they are based on counts over the total area covered by both control fields. The fourth 
column of Table 5 shows the difference between the second and third columns.

	The entries in the last column of Table 5 indicate that 
there is an excess number of objects with respect to what is seen in the control field 
when $r < 90$ arcsec. The statistical significance of the excess population 
in the $30 - 60$ and $60 - 90$ arcsec intervals is modest, and this is 
due to the narrow radial binning interval. The statistical significance can be 
increased by considering wider radial intervals. For example, between 30 and 90 arcsec 
the number of excess objects is $17 \pm 9$, while between 20 and 90 arcsec this number 
is $31 \pm 11$. We identify this excess population as globular clusters in Maffei 1; 
in \S 4 these numbers are used to compute S$_N$ for Maffei 1. 

\subsection{The $(J-H, H-K)$ TCD and the Near-Infrared Colors of Globular Clusters in Maffei 1}

	Foreground stars complicate efforts to identify 
individual clusters in Maffei 1. Nevertheless, the contrast between 
the cluster and foreground star populations is sufficient to allow 
limited conclusions to be drawn regarding the near-infrared colors of clusters in 
Maffei 1. Such an investigation is aided by the high extinction towards 
Maffei 1, which facilitates efforts to distinguish between foreground 
stars and globular clusters on the TCD. In particular, extragalactic sources will be 
the most heavily reddened objects along the line of sight, while foreground stars will be 
subject to a range of reddenings; as a result, the heavily reddened Maffei 1 clusters may 
have observed colors that are redder than those of most stars. Another factor that 
contributes to efforts to detect Maffei 1 clusters is that the intrinsic near-infrared 
SEDs of globular clusters differ from those of foreground disk stars, the vast 
majority of which are on the main sequence. K and M main sequence stars in the solar 
neighborhood have unreddened colors $J-H < 0.67$, while the K and M giants that dominate 
the near-infrared SEDs of globular clusters can have unreddened $J-H$ colors approaching 
1.0 (Bessell \& Brett 1988).

	The $(J-H, H-K)$ TCD is a convenient means of examining the broad-band 
near-infrared SEDs of Maffei 1 globular clusters. The left hand panel of Figure 
8 shows the $(J-H, H-K)$ TCD of objects with $K$ between 16.6 and 18.0 in Pointings 2, 
3, and 4 that are located between 20 and 90 arcsec from the center of Maffei 1. 
The composite TCD of the two control fields is shown in the middle panel; the number of 
objects in the middle panel greatly exceeds that in the left hand panel because the 
control fields sample a much larger area on the sky than the modest radial interval 
considered in Maffei 1. Also shown in Figure 8 is the reddening vector and the 
sequence defined by solar neighborhood K and M main sequence stars from 
Bessell \& Brett (1988). The majority of objects in the control fields have 
$J-H < 0.9$, which is what would be expected for K and M main sequence 
stars with A$_V \sim 2$ magnitudes.

	The TCDs in Figure 8 have been divided into $0.2 \times 0.2$ magnitude cells to 
facilitate the statistical analysis of the distribution of sources. 
The difference between the number of objects in each cell in the Maffei 1 TCD and the 
number in the corresponding cell in the control field TCD, with the latter scaled to the 
area searched for clusters in Maffei 1, is shown in the right hand panel of 
Figure 8. The number of objects in the Maffei 1 field exceeds that in the control field 
in all but one of the eight cells, and in three cells 
the difference is significant at roughly the $2-\sigma$ level or higher.

	The cells where there are significant excess populations, which are likely 
due to globular clusters in Maffei 1, span the full range of colors examined, with 
$J-H$ between 0.5 and 1.3, and $H-K$ between 0.2 and 0.6. 
Adopting A$_V = 4.7$, the color excesses computed from 
the Rieke \& Lebofsky (1985) reddening curve are $E(H-K) = 0.30$ 
and $E(J-H) = 0.50$. Thus, significant excess populations occur in cells with 
$(J-H)_0$ between 0.0 and 0.8 and $(H-K)_0$ between --0.1 and 0.3. For comparison, 
data from Barmby et al. (2000) and Barmby, Huchra, \& Brodie (2001) indicate that 
the majority of globular clusters in M31 have 
$(J-H)_0 = 0.6 \pm 0.2$ and $(H-K)_0 = 0.1 \pm 0.1$. The globular clusters in Maffei 1 
thus cover a range of intrinsic colors that is at least as large as that seen among 
M31 clusters. The clusters in the cell in the lower left hand corner of the 
TCD have infrared colors that are bluer than those of M31 clusters, and we 
note that 2 of the 20 cluster candidates identified by Buta \& McCall (2003) in 
Maffei 1 have $(R-I)_0 \sim 0$. While the nature of these objects is a matter of 
speculation, one possible interpretation is that they might be young clusters.

	The five objects in the cell with $J-H > 1.1$ and $H-K > 0.4$ are of particular 
interest, as no objects with this color are seen in the control fields. After correcting 
for extinction, the objects in this cell have intrinsic colors that fall within the range 
seen among members of the M31 cluster system, and so they do not have peculiar 
near-infrared SEDs. Given the absence of objects with similar colors in the control 
field, we expanded the search for objects in this cell to cover all of Pointings 
2, 3, and 4, and an additional 8 objects were found.

	The spatial distribution of these red objects 
provides a means of checking if they are truely clusters, as the density of 
clusters is expected to rise towards smaller galactocentric radii. The artificial star 
experiments indicate that the CFHTIR data are most sensitive to 
sources more than $\sim 30$ arcsec from the center of Maffei 1. 
Nevertheless, 5 of the 13 cluster candidates identified throughout 
the area imaged with CFHTIR are located within 
30 arcsec of the nucleus, suggesting that the cluster candidates are concentrated 
near the nucleus of Maffei 1. The spatial distribution of the 
red objects is examined in Figure 9, where the differential and integrated counts 
of these objects in 4 radial intervals are compared with the predictions for a 
uniformly distributed population; the number counts have been 
corrected for incompleteness. It is apparent that the majority of sources found on the 
TCD are concentrated towards the nucleus, with a statistical 
significance that is slightly in excess of the $2-\sigma$ level in the innermost bin. 
Therefore, we conclude that the spatial distribution of the very red objects on 
the TCD is consistent with them being globular clusters. 

	The main result of this section is that the clusters around Maffei 1 span a large 
range of near-infrared colors, and this can be checked by investigating the colors of 
globular clusters identified using independent criteria. Buta \& McCall (2003) used the 
non-stellar appearance of objects in WFPC2 images to identify clusters near the center 
of Maffei 1. The extended nature of an object is a criterion 
that is systematically different from that employed here, and should be
unbiased in terms of color. Five of the Buta \& McCall (2003) candidate clusters 
were detected in all three filters in the CFHTIR data, and 
the near-infrared brightnesses and colors of these objects are listed in Table 6. The 
$K$ magnitudes of these clusters place them at the bright end of the GCLF. One of 
these objects falls outside of the region investigated in Figure 8, while 3 of the 
remaining 4 clusters fall in regions of the near-infrared TCD where a statistically 
significant excess with respect to the control field counts were found. The clusters 
identified by Buta \& McCall (2003) thus span a wide range of near-infrared colors, in 
agreement with the analysis of Figure 8.

\subsection{Very Red Objects}

	There are a modest number of objects in the Maffei 1 and control fields with 
$J-H$ and $H-K$ colors that are much redder than the box limits marked in Figure 8. 
There are 7 such objects in Pointings 1 - 4, and 
3 such objects in the two control fields. The number density of these objects 
is therefore the same in the control and Maffei 1 fields, 
indicating that they are not associated with Maffei 1.
These objects tend to have only a modest object-to-object dispersion in 
$H-K$ and $J-H$ colors, with mean values $\overline{H-K} = 0.94 \pm 0.13$ and 
$\overline{J-H} = 1.30 \pm 0.15$; the quoted uncertainties are 
the standard deviations about the mean. 

	The projected density of the very red objects, which have 
$K$ between 16.8 and 17.8 with a mean near $K = 17.2$, is $\sim 500$ degree$^{-2}$. 
For comparison, the density of galaxies with $K_0 = 16.7$ 
is $\sim 800$ degree$^{-2}$ (0.5 magnitude)$^{-1}$ (e.g. McCracken et al. 2000), 
and so the majority of very red objects in our data are likely galaxies. 
The near-infrared colors of the very red objects are consistent with this 
claim. Fioc \& Rocca-Volmerange (1999) investigated the 
integrated near-infrared colors of nearby galaxies, and found that elliptical and spiral 
galaxies typically have $J-H \sim 0.75$ and $H-K \sim 0.25$. Now, if the very 
red objects are galaxies at intermediate redshift then 
k-corrections will be significant. In fact, Cowie et al. (1996) tabulated 
redshifts for galaxies in two fields, and the entries in their Table 1 indicate that 
galaxies with $K_0$ between 16.3 and 17.3 have a tight range of redshifts, with $<z> = 
0.33$ and a standard deviation $\pm 0.14$. Mannucci et al. (2001) computed k-corrections 
in the near-infrared, and the data in their Figure 7 indicate 
that for z = 0.3 the k-corrections in $J, H$, and $K$ are k$_{Corr}^{J} = 0.1$, 
k$_{Corr}^{H} = 0.0$, and k$_{Corr}^{K} = -0.6$ independent of morphology. Thus, a 
`typical' z = 0.3 galaxy when viewed through A$_V = 4.7$ magnitudes of extinction 
would have $J-H \sim 1.35$ and $H-K \sim 1.15$; these are in good agreement with the 
colors of the very red objects in the Maffei 1 and control fields.

	A modest number of the very red objects may also be L dwarfs. 
Kirkpatrick et al. (1999; 2000) find 62 L dwarfs to the 2MASS 
completeness limit of $K \sim 14.3$ in an area of 371 degree$^2$, and so 
the projected density on the sky is 0.2 degree$^{-2}$. 
Adjusting for the greater depth of our observations and assuming a uniform 
spatial distribution for L dwarfs in the solar neighborhood, the expected L dwarf 
density based on the Kirkpatrick et al. (2000) survey 
is 50 degree$^{-2}$ to our faint limit of $K = 18$, and so $\sim 1$ of the very red 
objects may be an L dwarf. In fact, the near-infrared SEDs of the very red objects 
are consistent with some fraction being L dwarfs. This is 
demonstrated in Figure 10, which shows the TCD of the L dwarfs discovered 
by Kirkpatrick et al. (1999, 2000) and the very red objects found here. If the objects 
in our data are L dwarfs then they would be intrinsically faint, and hence close to the 
Sun. Consequently, they will not be viewed through the 4.7 magnitudes 
of absorption in $V$ that affects Maffei 1, and the reddening correction will be 
smaller than that indicated by the reddening vector in Figure 10.

	The K-band spectra of L dwarfs show deep 
absorption bands due to H$_2$O and CO. These features are 
very wide, spanning wavelength intervals in excess of $0.1\mu$m 
(e.g. Figure 5 of Kirkpatrick et al. 1999). There are narrow-band 
filters that measure the strengths of both of 
these molecules in the $K-$band, and a narrow-band filter survey 
would provide a means of discriminating between L dwarfs and distant galaxies in the 
vicinity of Maffei 1.

\section{THE SPECIFIC CLUSTER FREQUENCY NEAR THE CENTER OF MAFFEI 1}

	The data are limited to $K = 18$, and so any clusters 
that are fainter than this are missed. The number of missing clusters 
can be estimated by assuming a shape for the GCLF. We adopt the M31 
$K-$band GCLF of Barmby et al. (2001). However, their LF is affected by incompleteness 
at the faint end. To correct for this, the LF was assumed to be symmetric about 
the GCLF peak, and the entries at the bright end of the LF 
were reflected about the peak to fill in the faint end.
Integrating the result indicates that a search for globular 
clusters in Maffei 1 that is limited to $K$ between 16.6 and 18.0 (M$_K$ between --12.0 
and --10.6) would detect only the brightest 27\% of the total cluster population.

	Given that the magnitude range considered in \S 3.3 includes 27\% of the cluster 
population, then the total number of clusters in Pointings 2, 3, and 4 located between 
20 and 90 arcsec of the center is $(31 \pm 11)/0.27 = 115 \pm 41$. Using the $V-$band 
light profile for Maffei 1 shown in Figure 24 of Buta \& McCall (1999), and not counting 
the area sampled with Pointing 1, we find that M$_V = -19.9$ between 20 and 90 arcsec 
of the nucleus if the distance modulus is $28.1 \pm 0.2$ and A$_V = 4.7 \pm 0.2$. 
Consequently, the specific cluster frequency near the center of Maffei 1, including the 
uncertainties in the distance modulus and foreground absorption, is S$_N = 1.3 \pm 0.6$. 
 
\section{DISCUSSION \& SUMMARY}

	Near-infrared images recorded with the CFHTIR imager have been used to search for 
globular clusters in the central $6 \times 6$ arcmin$^2$ of Maffei 1, which is one of the 
nearest large elliptical galaxies. The area surveyed includes more than half of 
the integrated light from the galaxy. The cluster content is investigated in a statistical 
manner by searching for an excess population of sources with respect to foreground disk 
stars. This is done near the center of the galaxy, where the projected concentration 
of clusters is greatest. The specific cluster frequency, S$_N$, is $1.3 \pm 0.6$ 
between 20 and 90 arcsec of the galaxy center. 

\subsection{Uncertainties in the Specific Cluster Frequency}

	The S$_N$ computed here holds for the inner regions of the galaxy, and is 
not a global value. In fact, S$_N$ changes with radius in elliptical galaxies, in the 
sense of becoming larger with increasing radius (e.g. Blakeslee \& Tonry 1995; Harris 
1991). Because of such a possible gradient in S$_N$, our local S$_N$ for Maffei 1 should 
be regarded as a lower limit to the global value, at least to the extent that the spatial 
distribution of clusters in Maffei 1 may be similar to that in the Coma and Virgo 
elliptical galaxies considered by Blakeslee \& Tonry (1995) and Harris (1991).

	The current observations sample only the bright end of the cluster population in 
Maffei 1, and to compute the total number of clusters the GCLF of M31 was adopted. 
Dynamical processes, which are expected to be most effective in the dense central regions 
of galaxies, can cause globular clusters to be disrupted over time. These processes will 
preferentially destroy low mass, extended clusters. The shape of the 
GCLF near the galaxy center may thus change over time (e.g. Vesperini 2000; 2001), and 
may vary from galaxy-to-galaxy depending on stellar density and the dynamical 
state of the cluster system. While we have no means of monitoring the extent of 
this evolution among clusters in Maffei 1, we note that the 
innermost regions of the galaxy were avoided when measuring $S_N$. 

	There are also practical problems inherent to studies of the central regions 
of galaxies. Spurious noise sources may masquerade as clusters near the nucleus of 
Maffei 1, and the artificial star experiments indicate that these flucuations affect 
the brightnesses of real objects in this region. In an effort to avoid these problems, we 
did not search for clusters in the central 20 arcsec of Maffei 1, 
where noise flucuations are clearly present (e.g. Figure 2). 

	Ideally, the impact of these effects could be monitored directly by 
investigating how S$_N$ changes with radius in Maffei 1. However, 
foreground star contamination limits our ability to do this.
In an effort to assess radial effects, S$_N$ was computed using only the candidate 
clusters detected between 30 and 90 arcsec of the galaxy center. We find that 
S$_N = 0.8 \pm 0.5$, which is not statistically different from 
S$_N$ computed for the region between 20 and 90 arcsec from the nucleus.

\subsection{The Colors of Clusters in Maffei 1}

	Buta \& McCall (2003) found that the candidate clusters in their sample tend 
to have $R-I$ colors that are redder, and have a larger scatter about the mean $R-I$ 
value, than expected if the Maffei 1 and Galactic globular cluster systems are similar. 
While our data confirm that the globular clusters around Maffei 1 span a broad range of 
near-infrared colors, there is no evidence that they are dominated by very red clusters. 
Indeed, the five clusters that we have detected from the 
Buta \& McCall (2003) sample have a broad range of near-infrared 
colors, with some falling near the blue end of what is seen among M31 clusters, and some 
falling near the red end. This is consistent with the near-infrared colors of the larger 
population of sources detected between 20 and 90 arcsec of the center of Maffei 1 
(\S 3.4). We emphasize that the reddening towards Maffei 1 is 
uncertain (e.g. \S 1), and that the $R-I$ colors computed by Buta \& McCall (2003) are 
more prone to uncertainties in the reddening than the near-infrared colors measured here. 

\subsection{Comparisons With Other Systems}

	How does the S$_N$ of Maffei 1 compare with 
what is seen other elliptical galaxies? There is a well documented broad spread in S$_N$ 
values among elliptical galaxies. Harris (1991) lists S$_N$ for 27 elliptical galaxies, 
and the entries in his Table 1 range from S$_N = 0.4 \pm 0.3$ (NGC 3557) to $14 \pm 
3$ (NGC 4486). The mean S$_N$ in the Harris (1991) elliptical galaxy sample is 5.1, with 
a standard deviation of $\pm 3.5$. Kundu \& Whitmore (2001) measured S$_N$ in 28 
elliptical galaxies, and found that S$_N$ ranged between $0.5 \pm 0.1$ (NGC 
3379) to $17.4 \pm 22$ (NGC 5982), with a mean of $2.4 \pm 1.8$. 
The CFHTIR data indicate that S$_N$ near the center of Maffei 1 falls near the 
lower end of global S$_N$ values for elliptical galaxies. This suggests that Maffei 1 
likely experienced a star-forming history that is not atypical for elliptical galaxies, 
and that the SGMCs needed to form globular clusters evidently occured in 
comparatively modest numbers during the early history of the galaxy and any progenitor 
systems.

	The broad range in S$_N$ among elliptical galaxies is due in part to 
environmental effects, in the sense that the elliptical galaxies with the smallest S$_N$s 
tend to be those in low density environments (Harris 1991). 
Cen A (NGC 5128) is a well-studied elliptical galaxy that has a similar environment, 
distance, and bright red stellar content to Maffei 1 (e.g. Davidge 2002). However, 
unlike Maffei 1, Cen A contains a prominent dust lane and outer shell structures (e.g. 
Malin, Quinn, \& Graham 1983) that are signatures of a recent merger. The specific 
cluster frequency in Cen A is S$_N = 1.4 \pm 0.2$ (Harris, Harris, \& Geisler 2004). 
While we caution that the cluster sample used by Harris et al. (2004) is dominated by 
objects in the outer regions of the galaxy, as the dust lane that slices through the 
center of the galaxy makes it difficult to detect objects within a few arcmin of the 
nucleus, the global S$_N$ in NGC 5128 is in excellent agreement with what we have 
measured near the center of Maffei 1; the list of similar characteristics of NGC 5128 and 
Maffei 1 thus may include the specific frequency of globular clusters.

\subsection{Future Work}

	Clearly, it is of interest to expand the sample of clusters in Maffei 1 to 
larger radii, and to go deeper. Because of the high density of foreground stars, it will 
be helpful to use criteria such as morphology, radial velocity, and spectral 
characteristics -- especially the depths of absorption features that are sensitive to 
surface gravity -- to identify individual clusters out to 
large galactocentric radii. Work of this nature would most 
easily be done in the infrared, given the substantial line of site extinction towards 
Maffei 1. Deeper samples are of particular interest, since these will allow the turn-over 
in the GCLF, which is a standard candle, to be identified. Indeed, the various distance 
moduli listed in Table 1 span 0.8 dex, and the detection of the turn-over in the GCLF 
will help resolve any uncertainty in the true distance modulus of Maffei 1.

\parindent=0.0cm
\clearpage

\begin{table*}
\begin{center}
\begin{tabular}{lll}
\tableline\tableline
Method & $(m-M)_0$ & Reference \\
\tableline
Surface Brightness Flucuations & $28.1 \pm 0.25$ & Luppino \& Tonry (1993) \\
Brightest AGB Stars & $28.2 \pm 0.3$ & Davidge \& van den Bergh (2001) \\
Fundamental Plane/$D_n - \sigma$ & $27.4 \pm 0.2$ & Fingerhut et al. (2003) \\
\tableline
\end{tabular}
\end{center}
\caption{Distance Modulus Estimates for Maffei 1}
\end{table*}

\clearpage

\begin{table*}
\begin{center}
\begin{tabular}{lcl}
\tableline\tableline
Method & A$_V$ & Reference \\
\tableline
Integrated color & $5.1 \pm 0.2$ & Buta \& McCall (1983) \\
Color of AGB Stars & $4.5 \pm 0.8$ & Davidge (2002) \\
Mg$_2$ -- Color relation & $4.67 \pm 0.19$ & Fingerhut et al. (2003) \\
\tableline
\end{tabular}
\end{center}
\caption{Extinction Estimates for Maffei 1}
\end{table*}

\clearpage

\begin{table*}
\begin{center}
\begin{tabular}{lll}
\tableline\tableline
Field & RA (J2000) & Dec (J2000) \\
\# & & \\
\tableline
1 & 02:36:35.4 & $+59$:09:19 \\
2 & 02:36:35.4 & $+60$:09:19 \\
\tableline
\end{tabular}
\end{center}
\caption{Co-ordinates of the Control Fields}
\end{table*}

\clearpage

\begin{table*}
\begin{center}
\begin{tabular}{lccc}
\tableline\tableline
Pointings & $\Delta J$ & $\Delta H$ & $\Delta K$ \\
Compared & & & \\
\tableline
1 \& 2 & $0.017 \pm 0.008$ & $0.040 \pm 0.017$ & $0.012 \pm 0.020$ \\
3 \& 4 & $0.007 \pm 0.017$ & $-0.006 \pm 0.012$ & $0.048 \pm 0.014$ \\
\tableline
\end{tabular}
\end{center}
\caption{Comparison of Photometric Measurements}
\end{table*}

\clearpage

\begin{table*}
\begin{center}
\begin{tabular}{lccc}
\tableline\tableline
Radial Interval & N$_{Maffei1}$ & N$_{Control}$ & $\Delta$N \\
(arcsec) & & & \\
\tableline
20 -- 30 & $20.0 \pm 6.0$ & $5.7 \pm 0.3$ & $14 \pm 6$ \\
30 -- 60 & $31.0 \pm 5.6$ & $23.0 \pm 1.2$ & $8 \pm 6$ \\
60 -- 90 & $47.0 \pm 6.9$ & $38.3 \pm 2.0$ & $9 \pm 7$ \\
90 -- 120 & $59.0 \pm 7.7$ & $53.7 \pm 2.9$ & $5 \pm 8$ \\
\tableline
\end{tabular}
\end{center}
\caption{The Excess Number of Objects with $K$ Between 16.6 and 18.0 in Three Radial Intervals}
\end{table*}

\clearpage

\begin{table*}
\begin{center}
\begin{tabular}{lccc}
\tableline\tableline
Cluster \# & $K$ & $J-H$ & $H-K$ \\
\tableline
1 & 16.467 & 1.167 & 0.824 \\
12 & 16.867 & 0.763 & 0.326 \\
14 & 16.289 & 0.798 & 0.225 \\
17 & 16.715 & 1.132 & 0.543 \\
18 & 17.605 & 0.929 & 0.188 \\
\tableline
\end{tabular}
\end{center}
\caption{Near-Infrared Photometry of Globular Cluster Candidates Identified by 
Buta \& McCall (2003)}
\end{table*}

\clearpage

\begin{figure}
\figurenum{1}
\epsscale{1.0}
\plotone{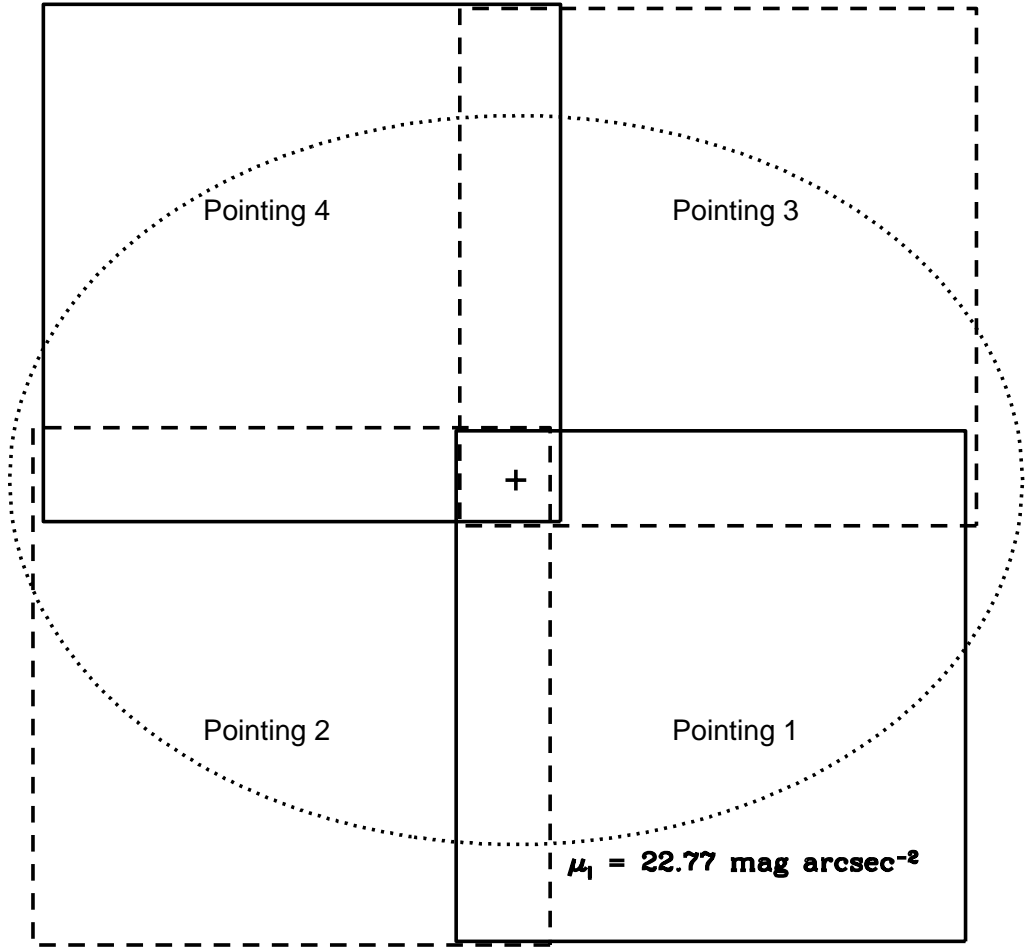}
\caption
{The geometry of the pointings used to map the central $6 \times 6$ 
arcmin$^2$ of Maffei 1. North is at the top, and East is to the left. The cross 
marks the center of Maffei 1, while the dotted line indicates the $\mu_I = 22.77$ 
mag arcsec$^{-2}$ isophote, which contains half of the integrated light from 
Maffei 1 (Buta \& McCall 1999).}
\end{figure}

\clearpage

\begin{figure}
\figurenum{2}
\epsscale{1.0}
\plotone{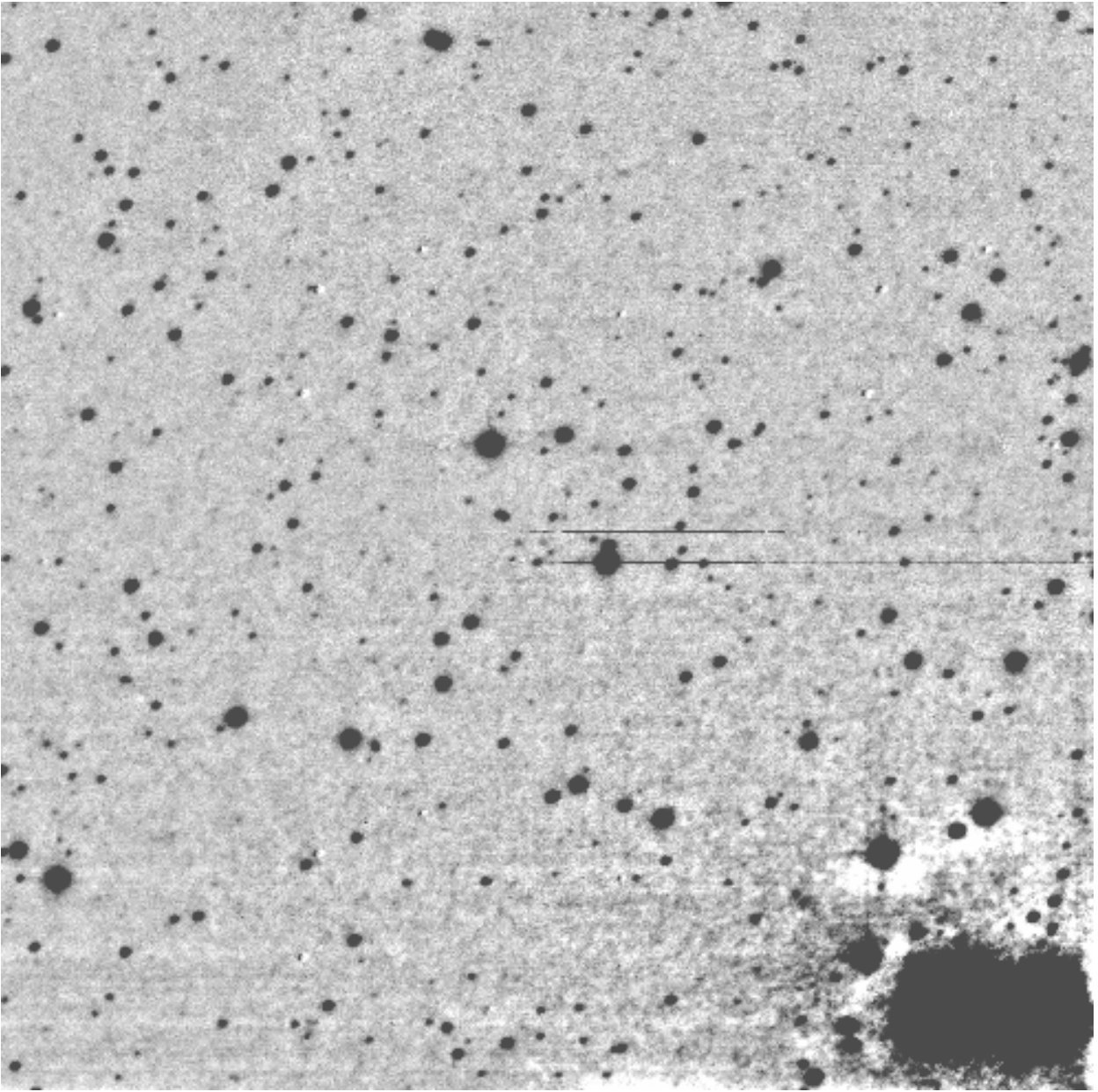}
\caption
{The final processed $K'$ image of Pointing 4, which covers $3.6 \times 3.6$ 
arcmin$^2$. North is at the top, and East is to the left. The 
center of the galaxy is in the lower right hand corner. The unresolved body 
of Maffei 1 has been removed using the procedure discussed in Section 3.1. The majority 
of objects in this image are foreground stars.}
\end{figure}

\clearpage

\begin{figure}
\figurenum{3}
\epsscale{1.0}
\plotone{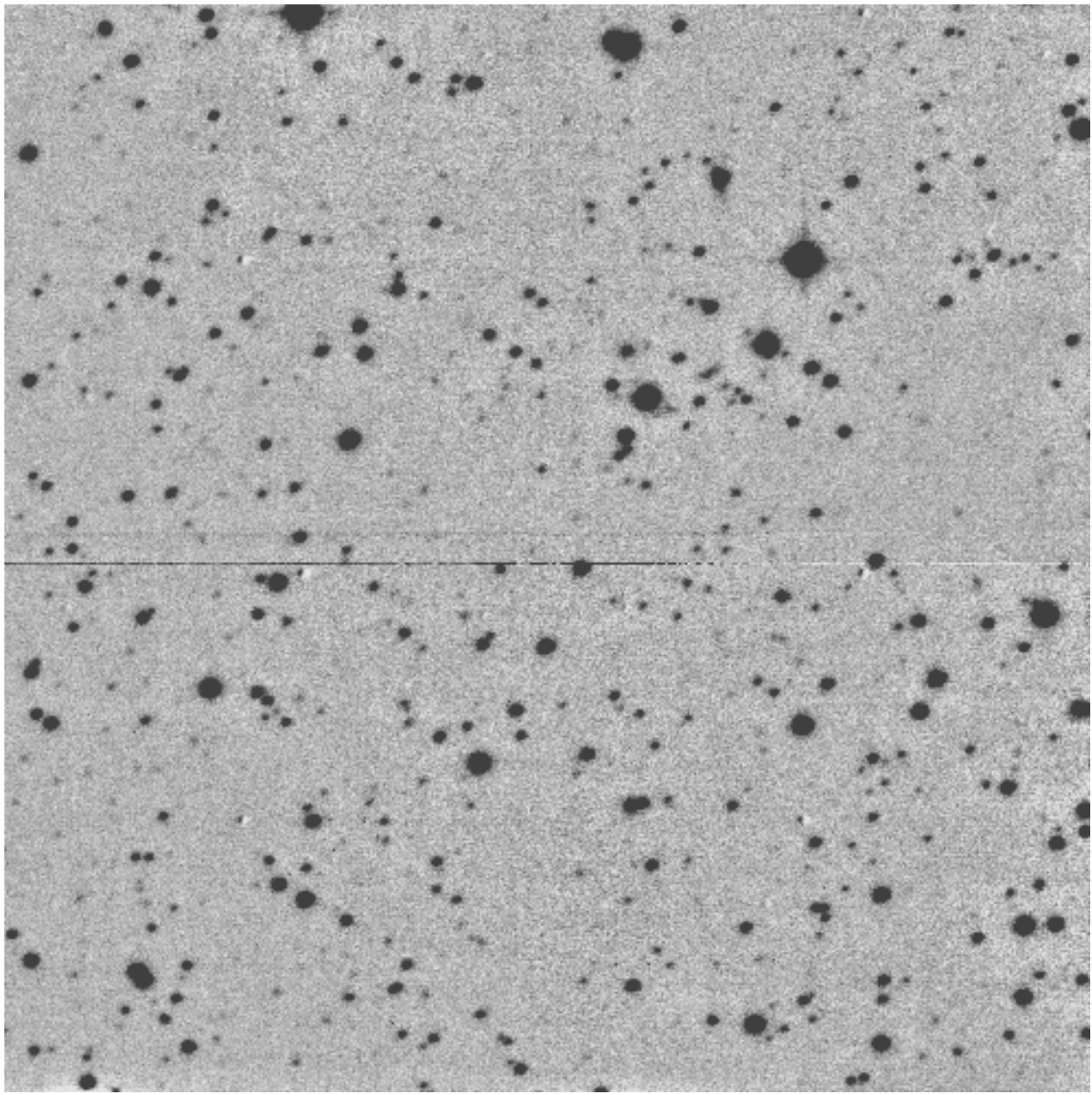}
\caption
{The final processed $K'$ image of Control Field 2, which covers $3.6 \times 
3.6$ arcmin$^2$. North is at the top, and East is to the left. Note that the density of 
objects in this field is roughly comparable to that in Figure 2, indicating that 
foreground disk stars dominate source counts near Maffei 1.}
\end{figure}

\clearpage

\begin{figure}
\figurenum{4}
\epsscale{0.9}
\plotone{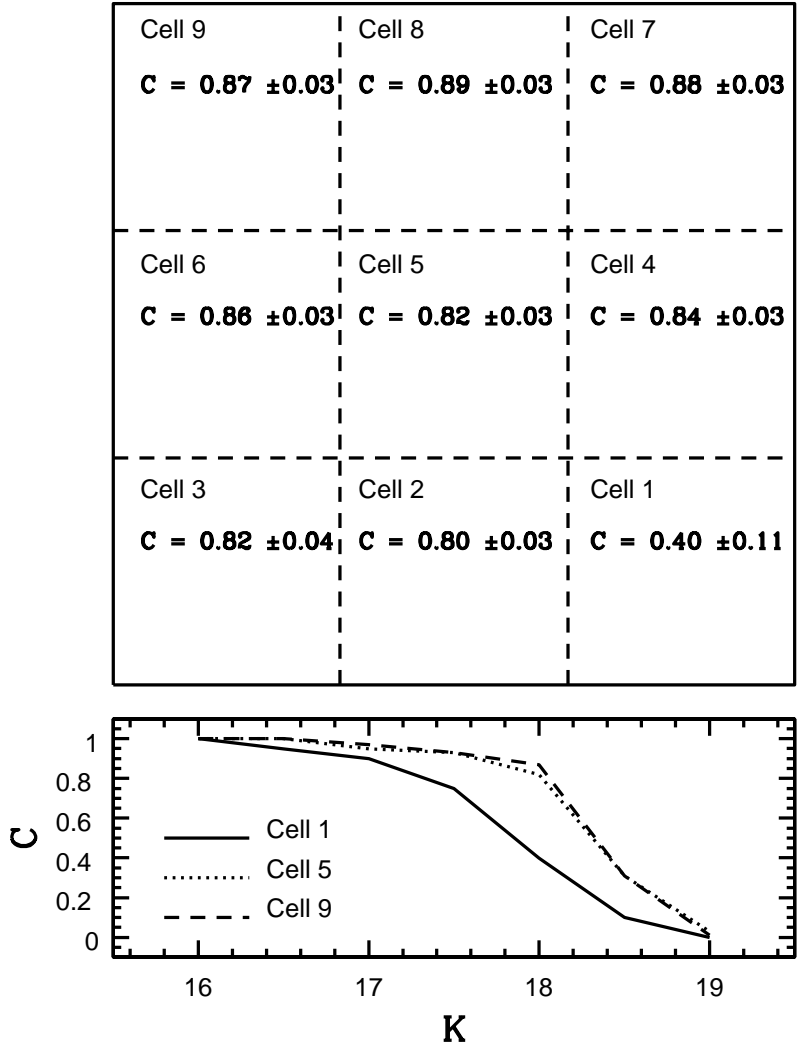}
\caption
{The behaviour of the completeness fraction $C$, which is the number of artificial stars 
recovered in all three filters divided by the total number that were added per magnitude 
interval, as a function of location and $K$ magnitude in Pointing 4. 
The top panel shows the spatial distribution of $C$ for sources 
in Pointing 4 with $K$ between 17.75 and 18.25. The cells have equal area, and are 
roughly 1.2 arcmin on a side. The center of Maffei 1 is located in Cell 1, and the 
elevated noise level resulting from the high surface brightness central regions of 
Maffei 1 causes $C$ in Cell 1 to be lower than elsewhere in Pointing 4. The lower panel 
shows $C$ as a function of $K$ in Cells 1, 5, and 9, which cut diagonally across 
Pointing 4. Note that the completeness fraction drops rapidly when $K > 18$ in 
Cells 5 and 9.}
\end{figure}

\clearpage

\begin{figure}
\figurenum{5}
\epsscale{1.0}
\plotone{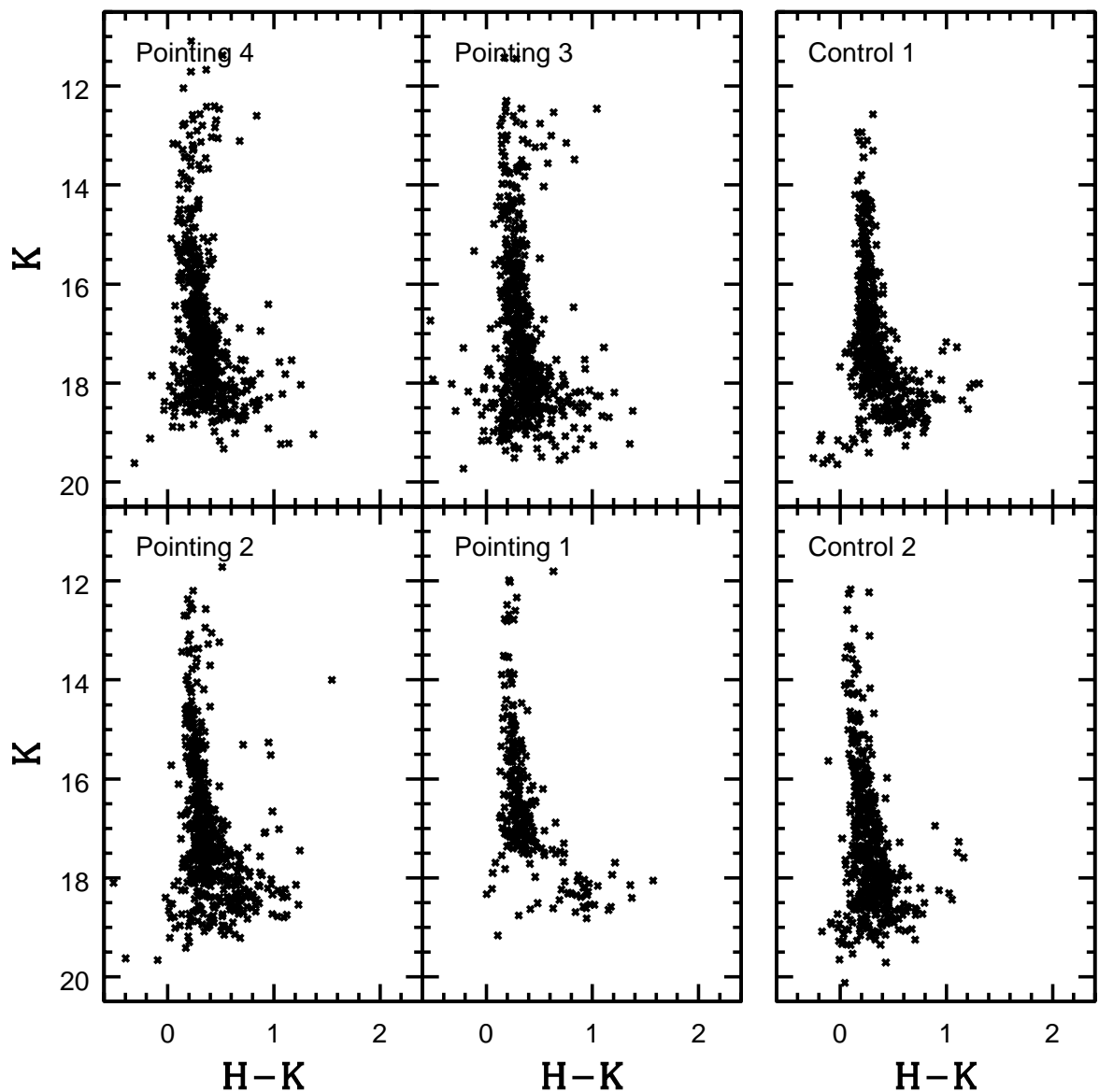}
\caption
{The $(K, H-K)$ CMDs of the Maffei 1 and control fields. The Maffei 1 CMDs are 
positioned according to the geometry indicated in Figure 1. Note the 
similarities between the Maffei 1 and control field CMDs, indicating that the vast 
majority of objects with $K < 18$ are stars in the foreground disk.}
\end{figure}

\clearpage

\begin{figure}
\figurenum{6}
\epsscale{1.0}
\plotone{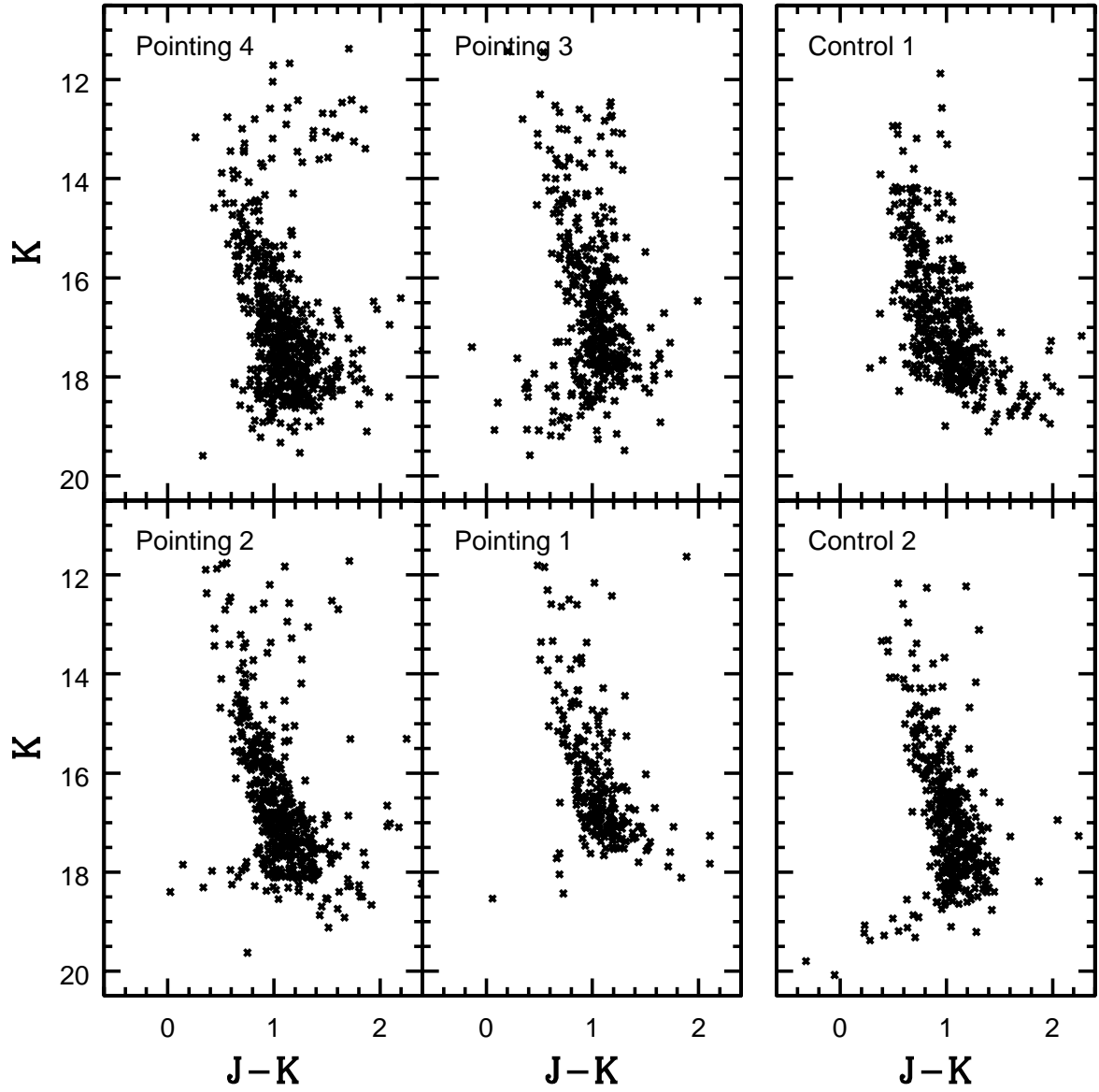}
\caption
{The same as Figure 5, but for the $(K, J-K)$ CMDs.}
\end{figure}

\clearpage

\begin{figure}
\figurenum{7}
\epsscale{1.0}
\plotone{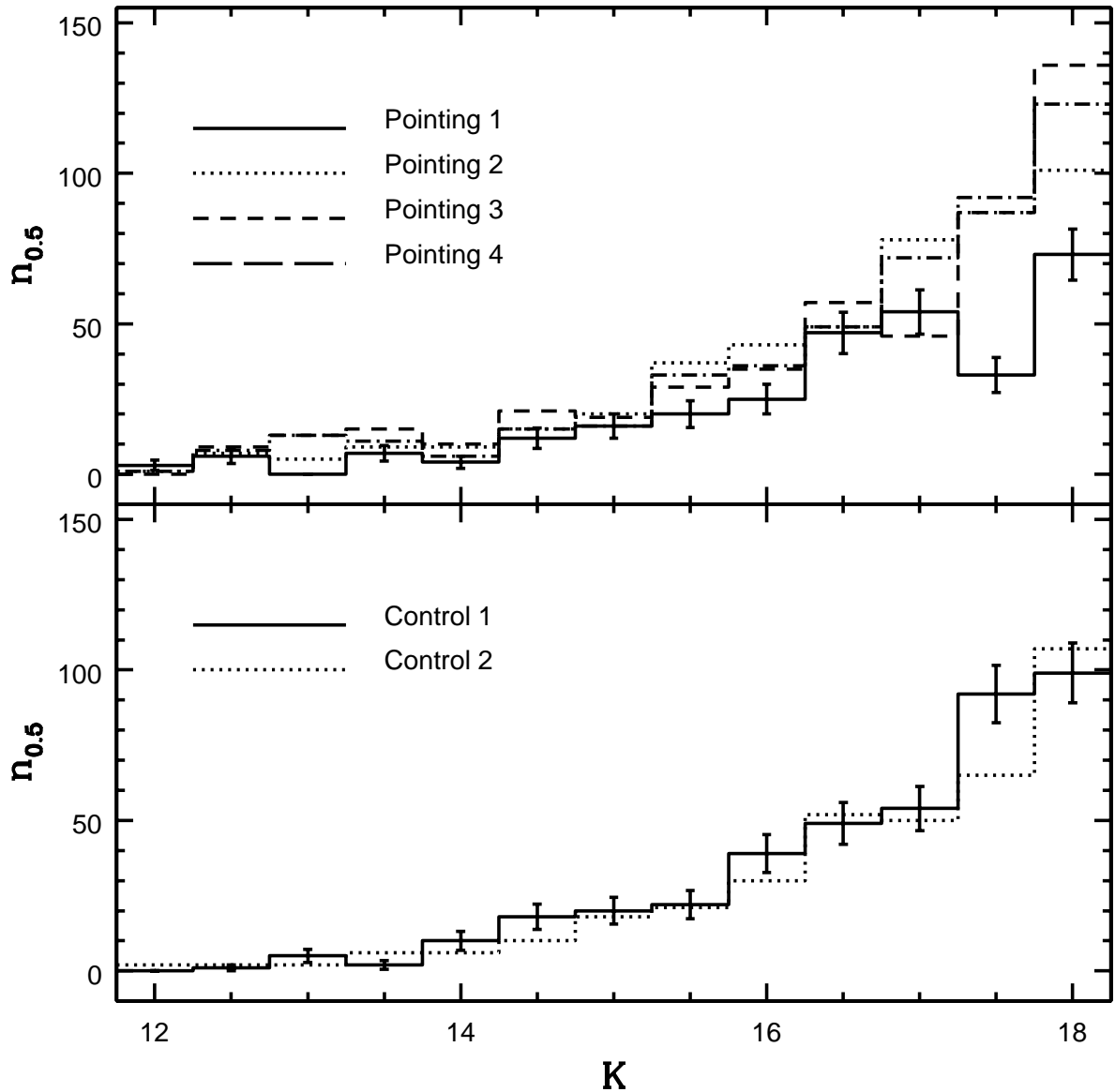}
\caption
{The $K$ LFs of the Maffei 1 and control fields. The error bars show the 
uncertainties in the Pointing 1 (top panel) and Control 1 (lower panel) number counts. 
n$_{0.5}$ is the number of objects per 0.5 magnitude interval in $K$, corrected for 
incompleteness. There is excellent agreement between the number counts 
in the two control fields, indicating that the foreground stars are uniformly 
distributed over 1 degree angular scales in this part of the sky. Note that 
Pointings 3 and 4 appear to have an excess number of sources near $K = 18$ when 
compared with the control fields; these excess sources are globular clusters in Maffei 1.}
\end{figure}

\clearpage

\begin{figure}
\figurenum{8}
\epsscale{0.78}
\plotone{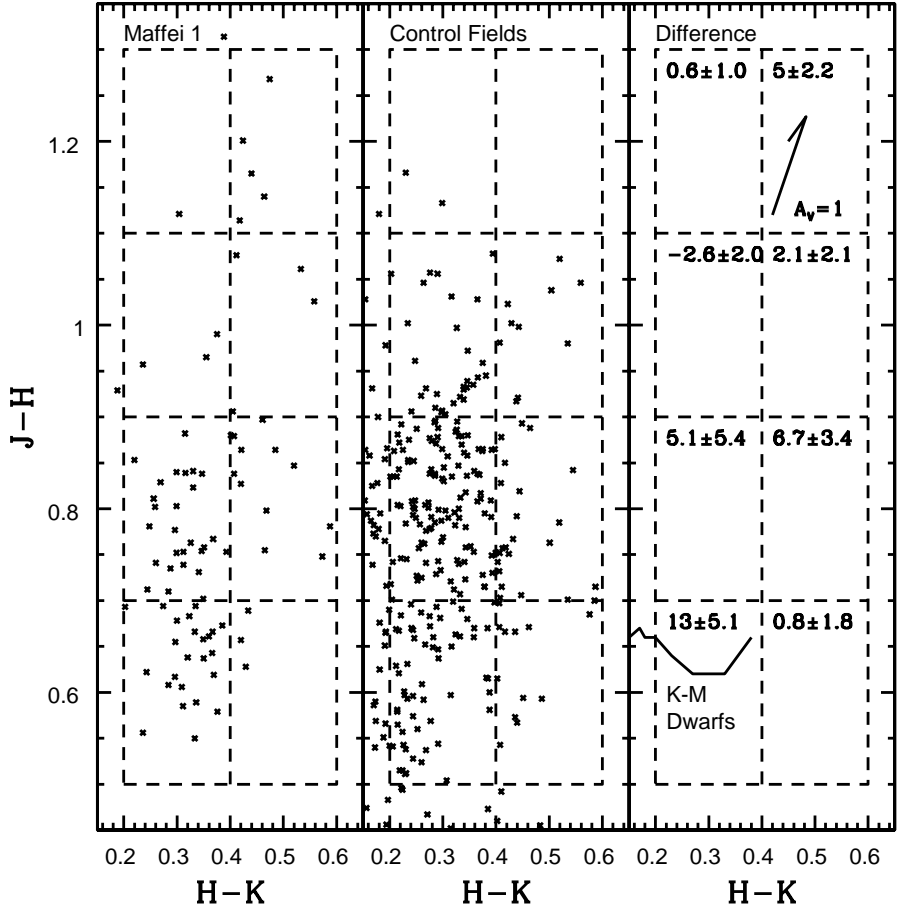}
\caption
{The left hand panel shows the $(J-H, H-K)$ TCD of objects with $K$ between 16.6 and 18.0 
that are located between 20 and 90 arcsec from the center of Maffei 1 in Pointings 2, 3, 
and 4, while the middle panel shows the TCD of objects with $K$ between 16.6 
and 18.0 in both control fields. The dashed lines in each panel define 
the $0.2 \times 0.2$ magnitude cells that are used to investigate the distribution of 
objects on the TCD. The entries in the right hand panel show the difference 
between the number of objects in the corresponding cell in the left hand 
panel and the number in the middle panel, with the latter scaled to the area in 
Maffei 1 that was searched for clusters. An excess population of sources 
is present at roughly the $2-\sigma$ significance level in three cells that span a 
range of $H-K$ and $J-H$ colors. A reddening vector with a length corresponding to 
A$_V = 1$ magnitude is also shown in the right hand panel, as is the 
sequence for solar neighborhood K - M main sequence stars from Bessell \& Brett (1988).}
\end{figure}

\clearpage

\begin{figure}
\figurenum{9}
\epsscale{0.9}
\plotone{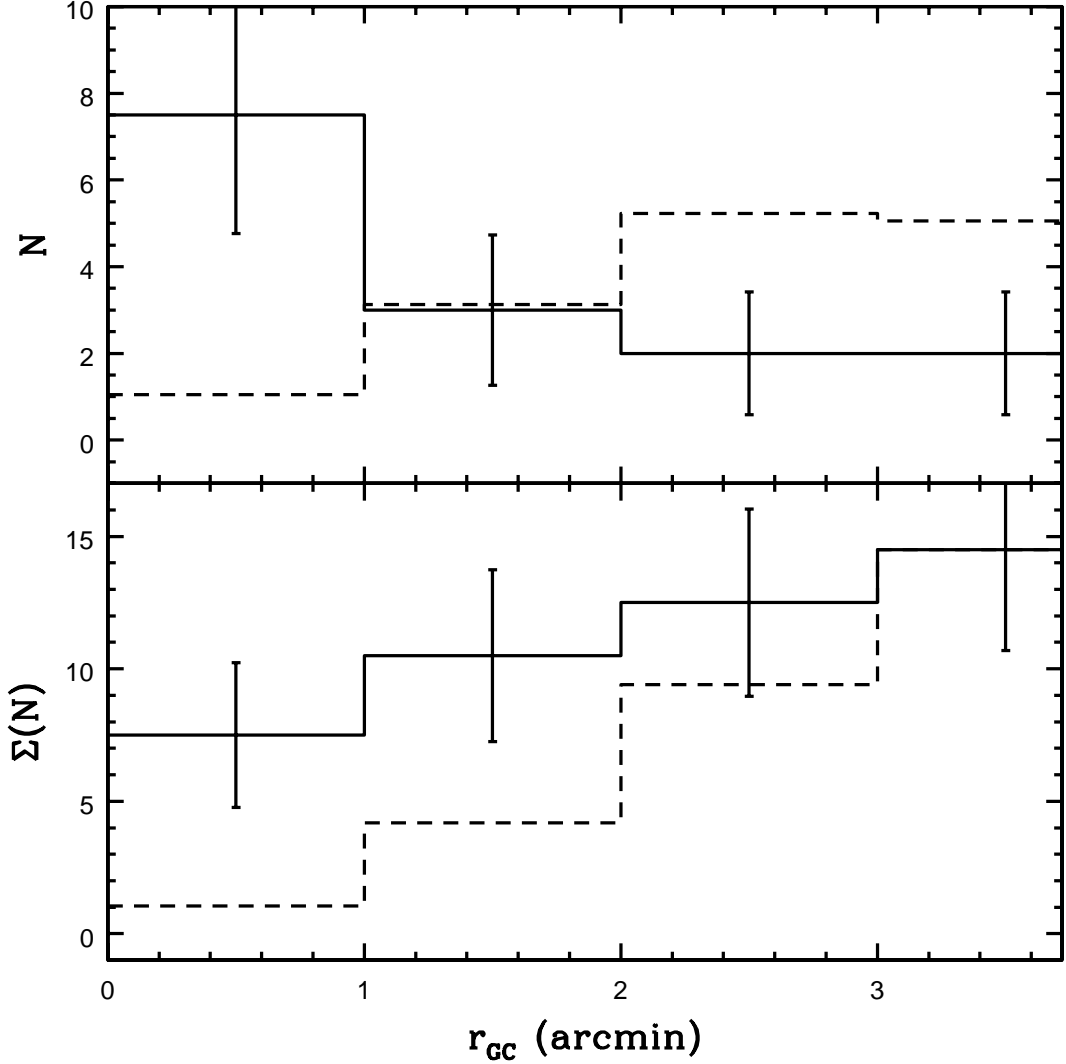}
\caption
{The radial distribution of very red objects detected in Maffei 1. The solid line in the 
top panel shows the number of cluster candidates with $H-K$ between 
0.4 and 0.6, and $J-H$ between 1.1 and 1.3 in 1 
arcmin wide radial bins, centered on the nucleus of Maffei 1. The dashed line shows the 
number of objects predicted per radial interval assuming that they are 
uniformly distributed. The solid line in the lower panel shows the cumulative number of 
cluster candidates in the same four radial bins, summing outwards to larger radii, 
while the dashed line again shows the trend for a uniformly distributed population. 
It is evident that the red objects are not uniformly distributed, but are concentrated 
near the center of Maffei 1, as expected if they are globular clusters.} 
\end{figure}

\clearpage

\begin{figure}
\figurenum{10}
\epsscale{1.0}
\plotone{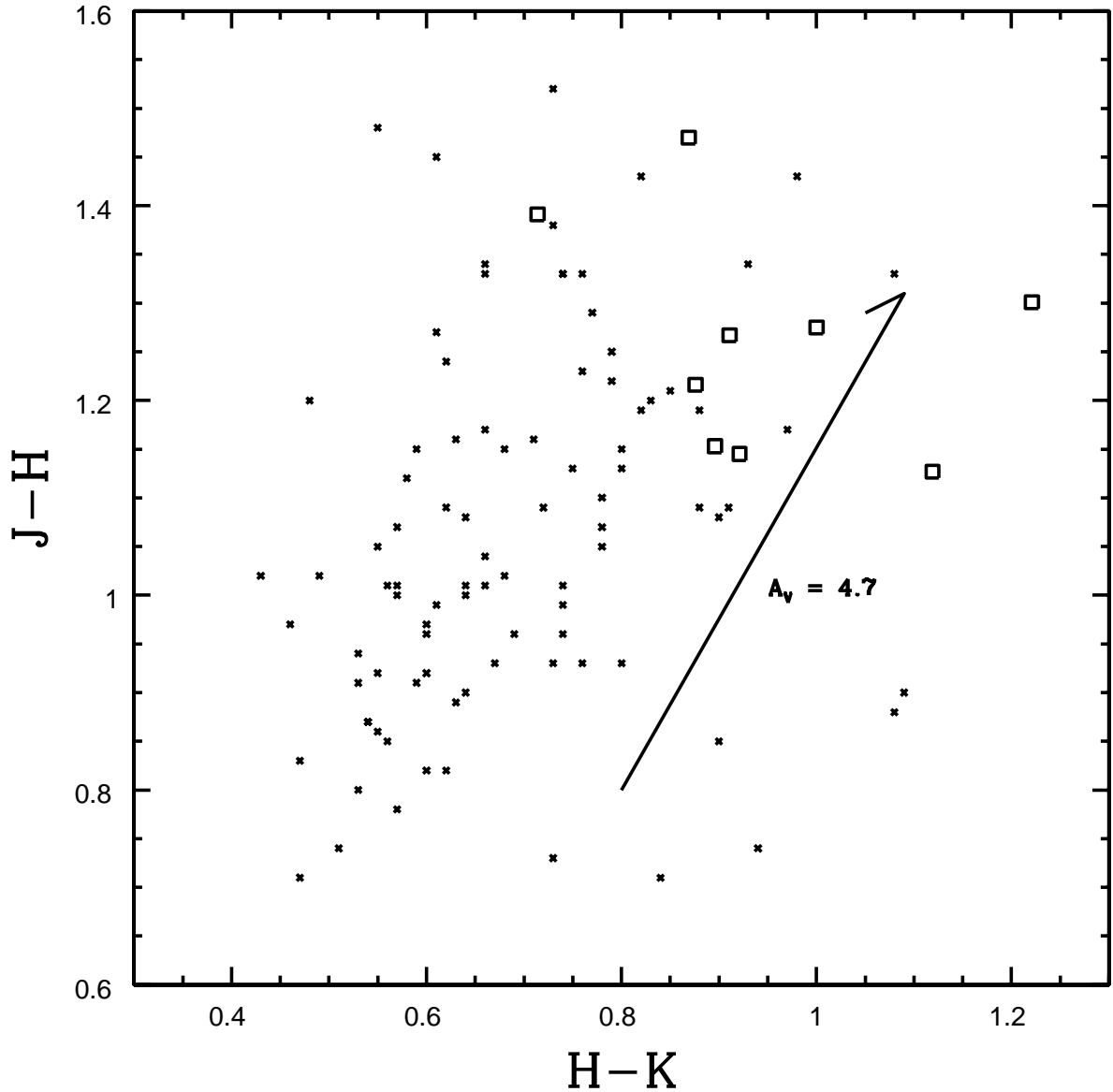}
\caption
{The $(J-H, H-K)$ TCD of spectroscopically confirmed L dwarfs (crosses) 
from Kirkpatrick et al. (1999, 2000) and the very red objects detected in our CFHTIR 
data (open squares). The arrow is the reddening vector, with an amplitude 
corresponding to A$_V = 4.7$. Note that the very red objects in the CFHTIR data have 
near-infrared SEDs that are similar to the majority of L dwarfs in the Kirkpatrick 
et al. (1999, 2000) sample.}
\end{figure}

\end{document}